# Reprogrammable and reconfigurable mechanical computing metastructures with stable and high-density memory

Yanbin Li[1], Shuangyue Yu[1], Haitao Qing[1], Yaoye Hong[1], Yao Zhao[1], Fangjie Qi, Hao Su[1*], Jie Yin[1*]

[1]Department of Mechanical and Aerospace Engineering, North Carolina State University, Raleigh, NC, 27606, USA.

*Corresponding authors: hsu4@ncsu.edu, jyin8@ncsu.edu

**Abstract**: Previous mechanical meta-structures used for mechanical memory storage, computing and information processing are severely constrained by low information density and/or non-robust structural stiffness to stably protect the maintained information. To address these challenges, we proposed a novel reprogrammable multifunctional mechanical metastructure made by an unprecedented building block based on kinematic mechanism. The proposed meta-structure can achieve all abovementioned functionalities accompanying with high information density and promising structural stability. We attribute all these merits to the intrinsic kinematic bifurcations of structural units, which enable the periodic meta-structure with additional and independently deformable bi-stable structural segments, and multi-layered deformed configurations to significantly enlarge the available information bits. We validate the stable information storage are originated from the compatible deformations of local structural segments before and after bifurcations. We illustrated the stored information can be feasibly reprogrammed by magnetic poles. Our design strategy paves new way for creating novel functional mechanical metastuctures.

**INTRODUCTION**

Mechanical computing has recently emerged as a new strategy for information processing and storage by using novel mechanical systems to augment the traditional electronic computing (*1*). Different from the electronic form, information in a mechanical system can be particularly encoded in its deformed patterns enabled by the unique constituent materials properties (*2-7*) and/or structural forms (*8-19*). Recent advances in novel mechanical systems, especially multistable systems in the forms of mechanical metamaterials (*8*, *20-28*), origami/kirigami structures (*3*, *29-31*), and mechanical mechanisms (*9*, *32-34*) provide unconventional platforms for robust mechanical memory storage (*23*, *31*), information interaction and encryption (*24*), and mechanical computation (*10*, *17*, *21*, *26*, *29*). Specially, multistable mechanical systems allows their stored information safely interacting with and robustly adapting to external changing environments (*23*). A bistable mechanical unit has two stable states in different configurations, which represent mechanical binary digits (bits) in a '0' or '1' state for information reading and storage. Basic structural forms of bistable units include constrained beams, curved plates, dome shells, deformable origami/kirigami structures, and balloons (*35*). Periodic tessellation of the bistable units in 1D (*3*, *8*, *30*), 2D (*20*, *21*, *23*, *24*), and 3D (*36*, *37*) forms a multistable mechanical computing system with exponentially increased stable states, which provides a tremendous space for processing information bits.

Although recent studies in multistable structures have shown promising potentials in storing and processing binary information, there remains substantial limitations and challenges in mechanical computing systems for achieving analogous functionalities to their electronic forms. First, it is challenging to achieve stable memory storage in most bistable units (*4*, *17*, *23*, *30*). On the one hand, binary information in mechanical systems should be very easily manipulated by reversibly switching and snapping between two stable states. This corresponds to writing and erasing information in response to mechanical forces or external stimuli such as heat, light, electricity, moisture, and magnetic field. Upon removal of actuations, the bits structures stay in their stable states to retain the information without additional energy input. On the other hand, once written or erased, stable memory for information storage requires the binary states should not be easily interrupted (changed or switched to another state) regardless of external perturbations such as mechanical loading or extreme external stimuli. This is challenging for most current designs since they could be easily snapped back upon applying external loading (*2-4*, *21*, *24*, *29-31*). Second, previous designs are hard to (re)program at the single bit level, i.e., an individual bistable unit, because the deformation in the unit could be coupled with its neighboring units due to the deformation compatibility (*17*, *23*). Recent study on tileable mechanical computing metamaterials shows the promising reprogrammability with stable memory at the unit-cell level (*23*). It uses magnetic actuation to reversibly and independently switch each binary element made of a bistable shell (*23*). Third, most studies are limited to binary information processing (*2-4*, *7*, *20*, *21*, *23*, *26*, *28-30*). How to increase the information densities beyond binary information by reprogramming binary states to more states remains challenging and unsolved.

Here, we report mechanism-based multistable multifunctional mechanical computing metastructure with stable and high-density mechanical memory and, high reprogrammability, and high information densities. The metastructure is constructed from periodic planar tessellation of reconfigurable rigid cube-based building blocks (**Fig. 1A, i**). Each building block is comprised of 2 by 2 unit cells (**Fig. 1B**). Cubes are bonded at their edges through elastic rotational hinges to form flexible closed-loop mechanisms in both building blocks and unit cells for shape reconfiguration (**Fig. 1B**). Uni-axially stretching the planar metastructure leads to both a bifurcated and multistable state with periodic corrugated surface features (**Fig. 1A, ii**). We find that all the ridged segments can be designed to be bistable under the applied pre-strain. Each bistable elements can act as an independent binary unit by reversibly popping up ('1' state) or down ('0' state) via snapping under pulling or pushing forces (**Fig. 1A, iii**, and **fig. S1**, **A** to **C**). Such physical binary elements can be used for combinatorial information writing (**Fig. 1C, i** to **iii**), erasing, reading, and encryption, as well as voxels for information display (**Fig. 1D**). The information can be stably stored by releasing the pre-strain (**Fig. 1A, iv**, and **fig. S1D**). Beyond binary units, the bifurcated metastructure can be further reprogrammed into multi-level step-wise pop-up structures for storing multidimensional information (**Fig. 1C, iv**).

To better understand the mechanical memory, we first investigate the reconfiguration kinematics of both the comprised unit cells and building blocks, as well as the bistable behavior in the building block through both modeling and experimental studies. Then, we explore the remote magnetic actuation of the periodic tessellated metastructure at the single bit level for binary and beyond-binary information processing and storage. Lastly, we explore the applications of the metastructure platform in information encryption and interpretation and mechanical computing as logic gates.

**RESULTS**

**Shape transformation in the unit cell**

As shown in **Fig. 1B** and **Fig. 2A**, the unit cell is composed of $2 \times 2$ sub-units (**Fig. 1B**, **i** to **ii,** see more details in **fig. S2-S3**). Each sub-unit is constructed by symmetrically connecting 4 rigid cubes (2×2) with four elastic line hinges (i.e., one connection among two adjacent cubes, see **Fig. 1B, iii**) into an over-constrained system. The prototypes were made by bonding the 3D printed rigid polymer cubes with ultra-adhesive plastic tapes (see Materials and Methods for details). The sub-unit can be treated as a 4R (4: number of rigid links, i.e., cubes; R: rotatable joints, i.e., hinges) rigid closed-loop kinematic mechanism (**Fig. 1B, iii and iv**). Similarly, the unit cell can be considered as a 4R closed-looped mechanism but flexible and hierarchical, which is composed of four flexible links (i.e., four sub-units) connected by four symmetrical line hinges, generating a two-fold structural symmetry (**Fig. 1B**, **ii**). The sub-unit can only deform as a chain-like configuration either along *x*- or *y*-axis (**Fig. 2A, i** and Supplementary video S1). However, the unit cell can undergo deformation with both in-plane expansion and out-of-plane extrusion (Supplementary video S1), forming an internal structural loop in the center (marked by the black-colored circular arrow in **Fig. 2A**, ②).

The unit cell exhibits rich shape transformations with over 10 multi-story configurations by following three different paths as shown in **Fig. 2A**, i.e., Path 1 simply reconfiguring as a chain-like mechanism (**Fig. 2A**, **i**), and the other two similar Path 2 and Path 3 as a looped rigid mechanism (**Fig. 2A**, **ii** and Supplementary Video S1). Due to the structural symmetry, the reconfigured shapes in Path 2 and Path 3 show a mirror symmetry about *xy* plane (**Fig. 2A**). All the shape changes can be characterized by the two rotation angles ($\theta_x$, $\theta_y$) at the boundary hinges along *x*- and *y*-axis, respectively (see details in **Fig. 2A**, ②). **Fig. 2B** shows the corresponding angle relationship, i.e., kinematics, during shape changes in **Fig. 2A**, where all the shape transformation follows simple linear relationship. Specially, Fig. **2B** shows two branched points at ($\theta_x$, $\theta_y$) = (45°, 45°) (i.e., configuration ② in **Fig. 2A**) and ($\theta_x$, $\theta_y$) = (90°, 90°) (i.e., configuration ③ in **Fig. 2A**), which represent kinematic bifurcation points with their angles predicted by the Denavit-Hartbenberg rule (*38-40*) (see analysis details in **fig. S4-S6** and more details in Supporting Information). We note that during transforming from configuration ① to ③ along the Path 2-1, i.e., $\theta_x$ increases from 0° to 90°, **Fig. 2C** shows that the nominal in-plane side length W (see the inset of **Fig. 2C**) of the unit cell first increases nonlinearly to the maximum at the bifurcated configuration ② (i.e., $\theta_x = 45°$), and then symmetrically decreases to the original length, whereas the out-of-plane structural height H (see the inset of **Fig. 2C**) keeps increasing monotonically (**Fig. 2C**). As demonstrated later, this unique in-plane expansion and contraction of the unit cell intrinsically provide the physical basis for achieving bistable deformation in the building block.

Next, we explore leveraging the rich transformed architectures for information processing. The unit cell has 16 cubes. We project the height of each cube in the transformed multi-story architectures to a 2D 4 × 4 array to form an in-plane mapping height contour. Specially, we observe that for the bifurcated configuration ③ or its mirrored configuration ⑧, each cube in the unit cell has a unique mapping height varying from 1 (i.e., story 1) to 3 (i.e., story 3), as shown in **Fig. 2D**. When compared to its original uniform mapping height of 1 before transformation, it gives a height difference changing from 0, 1, and to 2 in each cube, which can be assigned a respective '0', '1', and '2' state for potentially increasing information density beyond binary as discussed later. In contrast, most of the other transformed architectures do not possess unique one-to-one mapping height due to the cube stacking or overlapping (e.g., configuration ④-1, ④-2 and ⑥). Thus, we can simply use the unique in-plane mapping height contour to represent the transformed 3D architecture for information processing.

**Transformable and bifurcated building blocks with high-density information memory**

As shown in **Fig. 3A, i**, given the two-fold structural symmetry of the unit cell, we create the building block with four-fold structural symmetry by combining four unit cells with 16 line hinges. The hinges are symmetrically placed on the top and bottom surfaces to constrain the building block with a minimum number of structural degree of freedom (DOFs), as well as to preserve the reconfiguration paths in the unit cell. **Fig. 3B** shows the collection of transformable configurations of the building block. Similarly, its transformed shapes can be characterized by the combinatorial rotation angles ($\theta_x$, $\theta_y$) in the four unit cells. ($\theta_x$, $\theta_y$) in the unit cell now becomes ($\theta_c$, $\theta_y$) with $\theta_c$ and $\theta_y$ defined as the rotation angle at the center and the corner of the building block (see **Fig. 3B, i**), respectively. Similar to the Path 2 or Path 3 in the unit cell, as $\theta_c = \theta_y$ increase from 0° to 45°, the building block transforms into a corrugated configuration with five out-of-plane extruded segments, i.e., 4 in the corners and 1 in the center. At $\theta_c = \theta_y = 45°$ in each unit cell, it transforms into a bifurcation state (**Fig. 3B, i**).

Starting from the bifurcation state, we note that all the five extruded segments can deform independently and compatibly by either further "lifting up" or "pushing back" without mutual interferences. Thus, the building bock exhibits significantly enhanced reconfigurability through the different combinations of its five independently deformable structural segments (see details in Supplementary Video S2). To explicitly describe each reconfigured paths, we define the angle-sum pair ($\sum_4 \theta_{i\_c}$, $\sum_4 \theta_{k\_y}$) with $i$ and $k$ representing the unit cell number from 1 to 4. At the bifurcated state, we have ($\sum_4 \theta_{i\_c} = 180°$, $\sum_4 \theta_{k\_y} = 180°$). **Fig. 3B** shows 9 selected combinatorial transformed shapes using two different reconfiguration ways. One is to selectively further lifting up $m$ ($m$=1, 2, 3 or 4) corner structural parts while pushing back the central part with $\sum_4 \theta_{i\_c} = 0°$, e.g., configurations ①, ②, ③ and ④-1 on the left column. The other is to further lifting up both the central and $m$ corner structural parts with $\sum_4 \theta_{i\_c} = 360°$, e.g., configurations ④-2, ⑤, ⑥, ⑦ and ⑧ on the right column. The corresponding transition paths with simple linear angle relationships are shown in **Fig. 3C**. For example, configuration ① and ④-2 show only one elevated segment at the corner with ($\sum_4 \theta_{i\_c} = 90°$, $\sum_4 \theta_{k\_y} = 0°$) and in the center with ($\sum_4 \theta_{i\_c} = 360°$, $\sum_4 \theta_{k\_y} = 0°$), respectively, whereas all the central and corner segments are elevated in configuration ⑧ with ($\sum_4 \theta_{i\_c} = 360°$, $\sum_4 \theta_{k\_y} = 360°$).

Similarly, by following Path 3 of the unit cell in **Fig. 2A**, we can get mirrored transformed configurations of **Fig. 3B**. For example, **Fig. 3D, i to ii** show the shape of configuration ⑧ and its mirrored counterpart via Path 3 (e.g., configuration ⑨, see more details in **fig. S7**), respectively. We find that all the transformed configurations show the unique one-to-one mapping height contours in the 2D 8 × 8 array, see **Fig. 3D, iii** for configuration ⑧ and ⑨ for example. Compared to the unit cell, the building block could possess much higher information densities due to its rich reconfigurability. The building block can achieve 30 ($2 \times 2^4 - 2$) distinct transformed configurations and thus 30 2D mapping patterns for processing and storing information.

**Bistable building block**

In addition to the independent combinatorial shape transformation features, the building block can also achieve bistable deformation modes. As illustrated in **Fig. 4A, i**, uni-axially stretching the building block (e.g., along *x*-axis) generates a corrugated structure. Upon fixing the boundary parts as shown in the prototype of **Fig. 4A, ii**, the non-bifurcated corrugated configuration with $\theta_c = \theta_y < 45°$ under the pre-stretched state can maintain stable. **Fig. 4A, ii** shows an example of a stable pre-stretched configuration with $\theta_{c-1} = \theta_{y-1} = 25°$ (the sub-index 1 indicates the stable state 1) under pinned constraints. Then, vertically pulling the central part leads it to snap through to the other stable state with a pop-up central segment, see the configuration in **Fig. 4A, iii** with an elevated height of δh compared with the configuration in **Fig. 4A, ii**). $\theta_{c-2}$ in the stable state 2 increases to 65° with $\theta_{c\text{-}1} + \theta_{c\text{-}2} = 90°$ while $\theta_{y\text{-}2} = \theta_{y\text{-}1} = 25°$ (the sub-index 2 indicates the stable state 2) remain unchanged. Note that the bifurcated state with $\theta_{c\text{-}bifurcation} = 45°$ represents an unstable state as discussed next.

To experimentally examine the bistability, we use a fiber-enhanced ultra-adhesive tape as the line hinges to facilitate the strain energy stored in the hinges under the pre-stretched state (see Materials and Methods for details). **Fig. 4A, ii-iii** and Supplementary video S3 show the bistable demonstration by manually pulling and pushing the central segment under a fixed pre-stretched state by two pins. Through the displacement-control uniaxial mechanical test method, we further validate such bistable deformation by vertically compressing the sample from its second stable state back to the first stable state (see **fig. S8**). The measured force-displacement curve in **Fig. 4B** shows a sudden force drop that corresponds to the occurrence of snap-through instability, followed by a large negative force area, indicating the bistability.

We use both experiments and simplified modeling to understand the underlying mechanism of the observed bistability. Considering the expansion-contraction reconfiguration in the building block with free boundary in **Fig. 3B**, we hypothesize that the bistability relies on the incompatible reconfiguration kinematics between the central $\theta_c$ and corner opening angles $\theta_y$ of the unit cells induced by the fixed boundaries. To validate it, we measure the variation of $\theta_c$ as a function of $\theta_y$ under fixed boundaries and compare to the case of compatible path without any constraints. **Fig. 4C** and the inset show that for all the initial confined stable configurations with $(\theta_c)_1 = 15°$, 25° and 35° under different pre-stretched strains, the central opening angle (the red-colored lines) under boundary constraints is always smaller than that of the non-constrained compatible path (the cyan-colored line). Therefore, during the bistable switch, it always needs to overcome both the additional elastic compression and the incompatible out-of-plane deformation of the structural parts surrounded the central stretched area.

Specially, by assuming the torsional stiffness of all rotating hinges as $k$ and simplifying the compressed structure as a spring model (see details in **fig. S8D**), we theoretically compare the elastic energy $E$ stored in the structure under both compatible and incompatible kinematic paths (see details in Supplementary Materials). The normalized $E$-$\theta_y$ curves in **Fig. 4D** shows that the energy of the incompatible path (with $(\theta_c)_1$ selected around 25º) is always higher than that of the compatible path with a monotonically increasing $E$-$\theta_y$ curve. In contrast, the $E$-$\theta_y$ curve of the incompatible path exhibits two local minimums ($E_{min-1}$ and $E_{min-2}$), which correspond to the initial and snapped stable states with $E_{min-2} > E_{min-1} > 0$, respectively. The peak point with $E_{max}$ corresponds to the unstable state, which is located at the bifurcation state with an angle of 45º indicating the unstable structural characteristic.

**Reprogrammable mechanical metastructure for stable memory**

Periodically assembling $m \times n$ building blocks creates a reprogrammable multistable mechanical metastructure (see the representative 3 by 3 design in **Fig. 5A**). When the chain-like path 1 along both *x*- and *y*-axis are constrained by bonding the cubes (see the bonded cubes in the zoom-in part of **Fig. 5A**), the undeformed metastructure has only one DOF and deform by following Path 2 (or equivalently Path 3) of the unit cell shown in **Fig. 2A**, **ii**. Given the kinematic bifurcation and bistable deformation in each building block, we find that all the local out-of-plane extruded structural elements in an uni-axially pre-stretched metastructure can reversibly and independently pop up ('1' state) or down ('0' state) with bistabilty. Thus, an $m \times n$ metastructure can generate a number of $3\times2^{2mn}-1$ different combinatorial binary states for information (re)writing and (re)erasing with high information densities (**Fig. 5B**).

**Fig. 5C** shows three selected combinatorial binary states in a $2 \times 2$ metastructure prototype (see more details in Supplementary Video S4). In the prototype, we merged the repeating structural parts among units to save materials without affecting their transformation behaviors of both units and the periodic metastructure. The pre-stretched prototype is partitioned into $3 \times 3$ zones with its binary states represented by a $3 \times 3$ matrix **P** with $\mathbf{P} = \begin{bmatrix} A & B & A \\ B & A & B \\ A & B & A \end{bmatrix}$. A, B = '0' or '1' represents the binary state of flat or pop-up configuration in the segment, respectively. To write the information of $\mathbf{P_1} = \begin{bmatrix} 0 & 0 & 0 \\ 0 & 1 & 0 \\ 0 & 0 & 0 \end{bmatrix}$ into the metastructure, we follow two steps: first is to pull the center segment to snap and pop up ('1' state) under the fixed pre-stretched state by the constrained boundaries (see the process from **i** to **ii** in **Fig. 5A**); second is to release the pre-stretch to spontaneously flatten the corrugated units ('0' state) alongside lateral compression if needed without affecting the pop-up segment, forming a compact configuration (see the process from **ii** to **iii** in **Fig. 5A**). Similarly, we can write more binary information in a matrix form into the system with different combinatorial pop-ups/down. Furthermore, for the same number of pop-ups, e.g., $\mathbf{P_2} = \begin{bmatrix} 1 & 0 & 1 \\ 0 & 0 & 0 \\ 1 & 0 & 1 \end{bmatrix}$ with four evenly distributed pop-ups throughout the platform, we can also follow path 3 to write different information of $\mathbf{P} = \begin{bmatrix} 0 & 1 & 0 \\ 1 & 0 & 1 \\ 0 & 1 & 0 \end{bmatrix}$ with 4 centered pop-ups shown in **Fig. 5C** and **Fig. 5D**.

We find that the deformed structure with compact configuration is mechanically stable for storing information. We validate this by conducting compression test based on the pop-up configuration shown in **Fig. 5C**, **iii** with a single pixel (or equivalently information bit). **Fig. 5E** shows its compression force-displacement behavior without boundary constraints. It shows that it can bear about 2.8 kPa pressure at an applied compression strain of 2% without collapse.

Furthermore, given the reversible bistablity, we can either selectively or entirely erase the stored information in certain zones of the platform. For example, for selective erase, we can re-stretch the platform slightly to return to q stable state, where the platform becomes editable under the re-stretched state. Then, selectively pushing down the pop-ups erases the related stored information. Re-compressing the edited structure returns to the stable states to stably store the edited information without additional energy inputs (e.g., the reverse process from **iii** to **ii** in **Fig. 5A** and see more details in Supplementary Video S3 and S4). Similar to disk formatting, upon applying a relatively larger re-stretching strain close to the bifurcation strain as 0.5, all the stored information can be erased and the whole structural platform becomes flattened.

The information density can be enhanced by further reprogramming the same metastructure to a multilayer configuration. As shown in **Fig. 5F**, starting from the configuration with one pop-up in the center, we can continuously pull up the top segment to generate a five-story pyramid-like configuration with stepwise features (see **Fig. 5G** for the prototype) through a two-step process (see more details in Supplementary Materials). The five-story meta-structure also shows a unique one-to-one mapping height contour in a 2D layout, where each cube represents a single state of either '0', '1', '2', '3' or '4' depending on its height for quaternary information processing. The physical prototype stands stable and can tolerate certain external load for stable storage of mechanical memory. As the number of building blocks further increases, the metastructure can achieve much higher information storage density than previous works (*23*, *24*, *30*, *31*) through both the combinatorial independent pop-ups and the higher multi-story pyramid-like structures (**figs. S9-S10**).

**Information writing under magnetic actuation**

Based on the reprogrammable mechanical metastructure platform, in the following, we further explore its versatile applications in information storage, information encryption, and mechanical logic computation.

**Fig. 6A** shows the multistep process of generating a single central pop-up strut as a voxel in a 3×3 metastructure by following the aforementioned procedures of pre-stretching, selectively pop-up, and release and compression. The 3×3 metastructure render a 5×5 matrix. We note that for different pre-stretched strains $\varepsilon_y$, unlike the free-expansion without constrains, the side length of the constrained metastructure almost remains unchanged during the bistable switch of local structural segments (**fig. S11** and Supplementary Video S4). Moreover, **Fig. 6B** shows that for the fabricated 3×3 prototypes, the maximum pulling/pushing force $F_{max}$ to overcome the energy barrier and trigger the snapping in the local structural elements remains low, where $F_{max} \approx 1.65$ N, 0.85 N and 0.35 N for $\theta_c$ ($\varepsilon_y$) = 12.5º (4%), 22.5º (7.2%), and 42.5º (10%), respectively. For the case of $\theta_c = 42.5°$ (below a theoretical $\theta_{c\text{-}bifurcation} = 45°$), $F_{max} \approx 0.35$ N is even smaller than its self-weight force (~ 0.55 N), which means practically the pop-up strut cannot stably maintain its second stable state under self-weight. When considering the effect of unavoidable self-weight, we find that the critical angle $\theta_{c\_critical}$ for enabling bistability in the prototype is lower than the theoretical value of 45° with $\theta_{c\_critical} \sim 33°$.

Given the small $F_{max}$ in triggering the independent bistability in the prototype, we further explore using the simple remote magnetic field to actuate the reversible bistable switch (see Supplementary Video S5) in the metastructure as a mechanical memory storage device for information storage and display ( e.g., letters or images). **Fig. 6A** shows the schematic and experimental demonstration of writing four letters of 'NCSU' onto the 3 × 3 platform through the untethered actuation by using permanent magnets (see more details in Supplementary Materials). Thin-plate-shaped permanents are attached to the top surfaces of 9 black-colored local struts (see the right-bottom inset in **Fig. 6A**, **i**). Under remote magnetic field (see the right-top inset in **Fig. 6A**, **i,** and more details in **fig. S12**), the local struts can be independently and remotely pulled up or pushed down to stay in their stable positions without affecting their neighboring elements (see **figs. S13 to S14** and more details in Supplementary Materials). The insets show the storage information in a 5 × 5 binary matrix for each character (e.g., "N", "C", "S", "U" from **ii** to **v** correspondingly), e.g. "N" = $\begin{bmatrix} 1 & 0 & 0 & 0 & 1 \\ 0 & 1 & 0 & 0 & 0 \\ 1 & 0 & 1 & 0 & 1 \\ 0 & 0 & 0 & 1 & 0 \\ 1 & 0 & 0 & 0 & 1 \end{bmatrix}$ and "S" = $\begin{bmatrix} 0 & 0 & 1 & 0 & 1 \\ 0 & 1 & 0 & 0 & 0 \\ 0 & 0 & 1 & 0 & 0 \\ 0 & 0 & 0 & 1 & 0 \\ 1 & 0 & 1 & 0 & 0 \end{bmatrix}$. Similarly, more complex patterns or even images can be stored and displayed using the bistable voxels in a large size metastructure platform. For example, a schematic smile face on a 4 × 4 platform (**Fig. 1D, i**) and a wolf-head on a 30 × 27 platform (**Fig. 1D, ii**). Given the deformation independence of each local structural element, the same metastructure could act as a pluripotent mechanical platform for writing/re-writing and erasing for stable information storage (see Supplementary Video S6 and S7).

**Information encryption and decryption**

Next, we further exploit the metastructure as a structure-based information encryption (SIE) device by purposefully encoding distinct information onto this multistable system. **Fig. 7A** shows the working mechanism for the designed SIE system. First, considering each pop-up strut as one single vertex, we can encode the encrypted information onto the uniquely patterned pop-up struts in the form of vertices, lines, and/or polygons. Second, we can decipher these patterns with an external reading system (e.g., a screen display) for information decryption and display (see more details in **fig. S15A**). For example, we encrypt the word information of "Information" and "Encryption" word information using simple geometrical shapes such as a pop-up triangle for encoding and decoding processes shown in **Fig. 7B** and **fig. S15B**. First, we encode two different words of "Information" and "Encryption" into two different bistable deformed triangular shapes (**Fig. 7B**, **ii** and **iv**). Second, an electronic interpreting system with displacement-sensing diodes and Arduino circuit board (**Fig. 7B**, **i** and **iii**, and see more details in **fig. S15A**, right) will detect and transform these physical information into unique electric signals. Third, interpreted by our self-developed code, these electric signals can be captured by the control system and paired with their prescribed word information. Lastly, we can precept the encrypted information through direct display on a computer screen (see the desktop in **Fig. 7B**, **ii** and **iv**). As demonstrated in **Fig. 7B**, we encrypt two words "Information" and "Encryption" onto two symmetrical right-angle triangle patterns (see more details in Supplementary Video S8).

Remarkably, given the distinct spatial locations of each structural element and their independent combinations, we note a large volume information can be encoded into our proposed metastructure platform even with a small number of units. For example, for the simple 3×3 platform, we can encode a large number of $N_{info} = 9!-1 = 362879$ different informational contents (for a $n \times n$ platform, $N_{info} = n^2!-1$). Thus, compared to previous designs (*23, 24*), our proposed system shows superior advantages in terms of simple structural forms, easy fabrication and actuation, and superior informational encryption capability.

**Mechanical logic gates**

Lastly, we explore the metastructure as simple mechanical logic gates. **Figs. 7C-7D** demonstrate the achievement of both "OR" and "AND" logic gate operations by utilizing independent bistabilities in local elements. To facilitate the reading of output information (see more details in **fig. S16A** and Supplementary Materials), we use a supported height-adjustable flat plate on the top to cover a small region of the platform. Its initial state is set as an output of "0". When the plate is even and elevated, it outputs "1", otherwise "0" for the cases of either being tilted or lowered. The configurations of the top plate are determined by the pop-up ("1") or pop-down ("0") motions of three supports bonded to the bistable elements as inputs. A pyramid support denoted as $P_1$ is placed in the center with two other neighboring supports surrounded, e.g., cuboids of $S_1$ and $S_2$ and pyramids of $P_2$ and $P_3$ for the "OR" and "AND" logic gate, respectively.

**Fig. 7C** and **fig. S16B** shows that when $P_1$ is popped up and fixed, popping-up either $S_1$ or $S_2$ or combined as inputs leads to a stable and evenly elevated plate on the top as an output of "1" for an "OR" operation, because one point contact at $P_1$ alongside one plane contact at $S_1$ or $S_2$ will render a stable and even surface. For the case of "AND" logic gate shown in **Fig. 7D** and **fig. S16C**, three pyramid supports $P_i$ are flexible to pop up or down, providing the point contacts to support the top plate. Only when the plate is supported by three pop-up point contacts, i.e., $P_1 = P_2 = P_3 = $ "1", it will generate a stable and evenly elevated plate as an output of "1" for an "AND" operation.

We note that most of previous mechanical logic meta-structures are limited to 1D and 2D structural forms (*1-3*, *19*, *20*, *26*, *29*, *41-43*) while for the first time our design extends the structural form of the mechanical binary logic computation to 3D structural form. In **Figs. 7D-7E**, we demonstrate the logic operation in only one zone. Particularly, given the independent bistability of each local elements, we note such design principle can be readily applied to multiple zones for conducting a myriad of parallel mechanical binary operations on the same meta-structure platform (see details in **fig. S17**). Moreover, by altering the structural components as schematically illustrated in **fig. S18**, we can also conduct "NOR" and "NAND" binary logic computations in our designed platform.

**DISCUSSIONS**

In summary, we proposed a reprogrammable multistable mechanism-based metastructure composed of planar and compact tessellation of reconfigurable cube-based building blocks. We explored its potential as a pluripotent mechanical computing platform through both experimental testing and proof-of-concept demonstrations. Kinematic bifurcation in the building blocks enables combinatorial structural reconfigurability and bistability of the local elements for enhanced deformation re-programmability, where the bifurcated state corresponds to an unstable state. The bistability in the local elements can be manipulated independently without interference with their neighboring elements for information storage and computing at the single bit level under both contact-based mechanical forces and remote magnetic actuation. Leveraging both kinematic bifurcation and bistability characteristics, we demonstrated the potential applications in selectively information writing, erasing, and high-density stable memory storage, as well as information encryption and logic gates.

Given the scale independence of the kinematic designs, we envision that the design could scale down to a small scale for mechanical memory with robust and high stored information densities or other applications including MEMS, haptic devices, and reconfigurable meta-surfaces for acoustic wave guide, particle transports, and directional flow control (*44*, *45*) (see the representatives in **fig. S19**). Meanwhile, upscaling the designs to the meter scale could also find potential applications in deployable temporary buildings (**fig. S9B**).

## MATERIALS AND METHODS

**Fabrication and mechanical test of the HMMs**

The cube-shaped structural elements (overall dimension as 2cm × 2cm × 2cm and shell thickness as 1mm) of the hierarchical mechanism-based meta-structure (HMM) are printed by PolyJet 3D printing (Convex Object 260, Stratasys). And the rigid VeroWhite material (~0.68GPa) is used. Then ultra-adhesive plastic tapes (Scotch, 4198W-SIOC) are used to connect the cube structural elements together to form into the periodic HMM. To enable the structural units of the HMM performing local bi-stable deformations, the fiber-enhanced ultra-adhesive tape (Fiberglass Reinforced Tape, BOMEI PACK) are used to firstly assist the feasible stretching of the HMM into the deformed structural state before bifurcation, and then maintaining as the stretched state with some metal pins (Amazon Basics Push Pins Tacks) by fixing the fiber tapes onto substrate. The bi-stable deformation feature of the structural unit is tested with uniaxial compression by INSTRON (5945).

**Magnetic actuation of the bi-stable deformation of structural units on HMM**

To magnetically actuate the local structural units of the HMM from their first stable state to the second stable states, a cylinder shaped Neodymium Magnets (Diameter as 10cm and Magnetized through thickness; 5862K331 from McMASTER-CARR) is used to actively attract the thick-plate shaped passive Neodymium magnet (length × width × thickness as 3cm × 1cm × 0.2cm, magnetized through thickness; 7048T29 from McMASTER-CARR). The passive magnets are attached onto the central parts of the structural units with double sided tapes (Scotch Double Sided Tape, 137DM-2)

**Acknowledgements**

The authors acknowledge the funding support from National Science Foundation under award number CMMI-CAREER-2005374.

**Author contributions**

Y.L. and J.Y. proposed the idea. Y.L. conducted theoretical and numerical calculations. Y.L. and S.Y. designed and performed the experiments. Y.L., S.Y., H.S. and J.Y. wrote the paper. H.S. and J.Y. supervised the research.

**Competing interests**

The authors declare no competing interests.

**Figures**

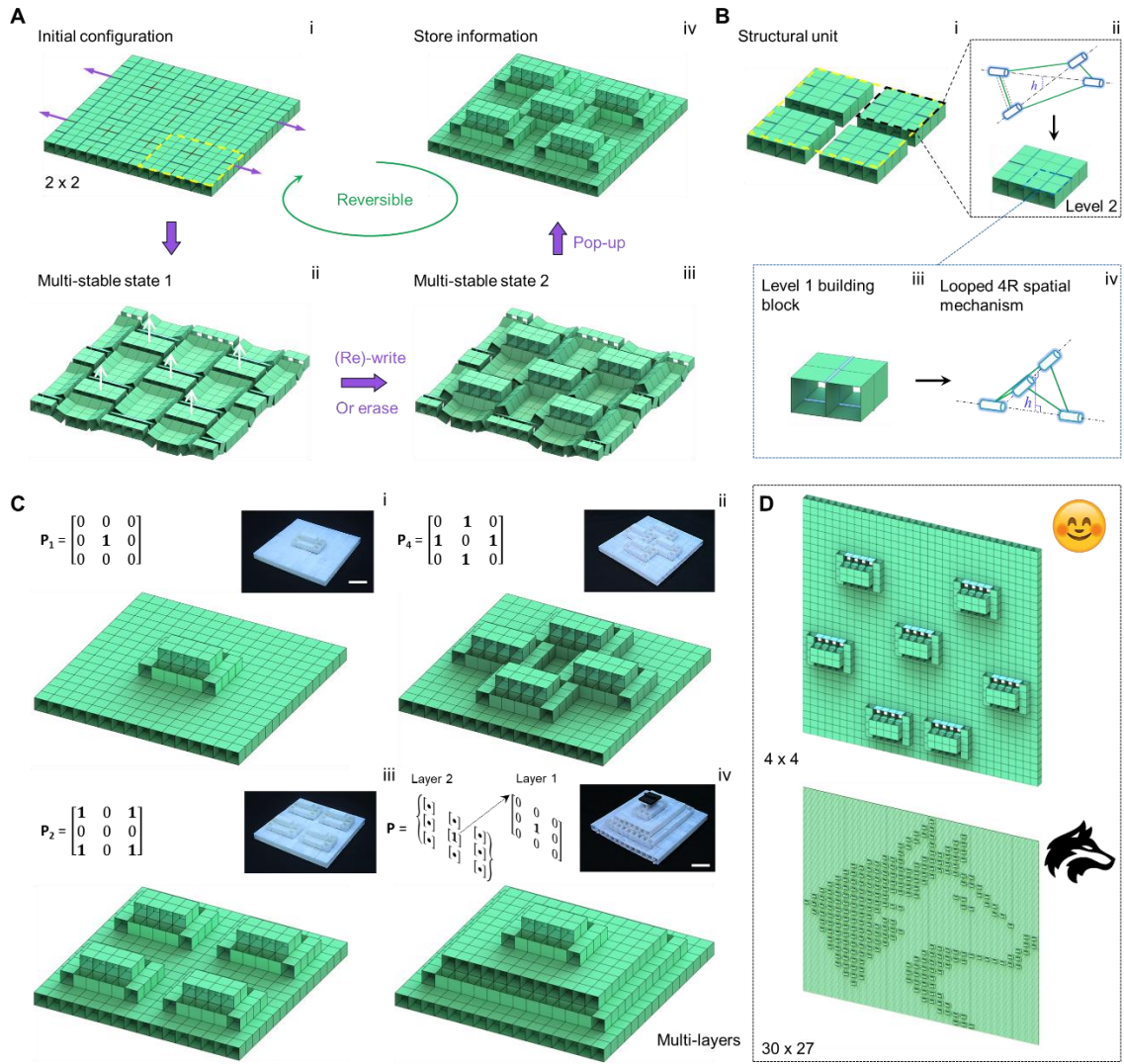

**Fig. 1. Schematics of the design and information patterning process of our proposed reprogrammable mechanical metastructure.** (**A**) The reversible information storage process (capable of erasing and rewriting) of the proposed mechanical metastructure with 2 by 2 building blocks from the initial configuration (i) to the first multistable state 1 (ii) to multistable state 2, and finally settling at the deformed compact configuration to store information stably (iv). (**B**) Construction process of the building block (i) based on structural hierarchy concept from four level-2 structural units connected by 4 line hinges (ii) while the structural unit is built by the level-1 4-cube based structure with 4 line hinges illustrated as (iii). (**C**) Demonstration of the mechanism of the high information density of our proposed mechanical metastructure through the reprogrammable combinatorial deformations of the independently reconfigured local structural segments (i to iii) and the multi-layered deformed configuration (iv). Scale bar: 2cm. (**D**) Illustration of the implementations on mechanical information storage and display on the 4 by 4 structure displayed with a smile face by seven local deformed segments (or equivalently the information bits) (i) and the 30 by 27 structure displayed with a wolf head.

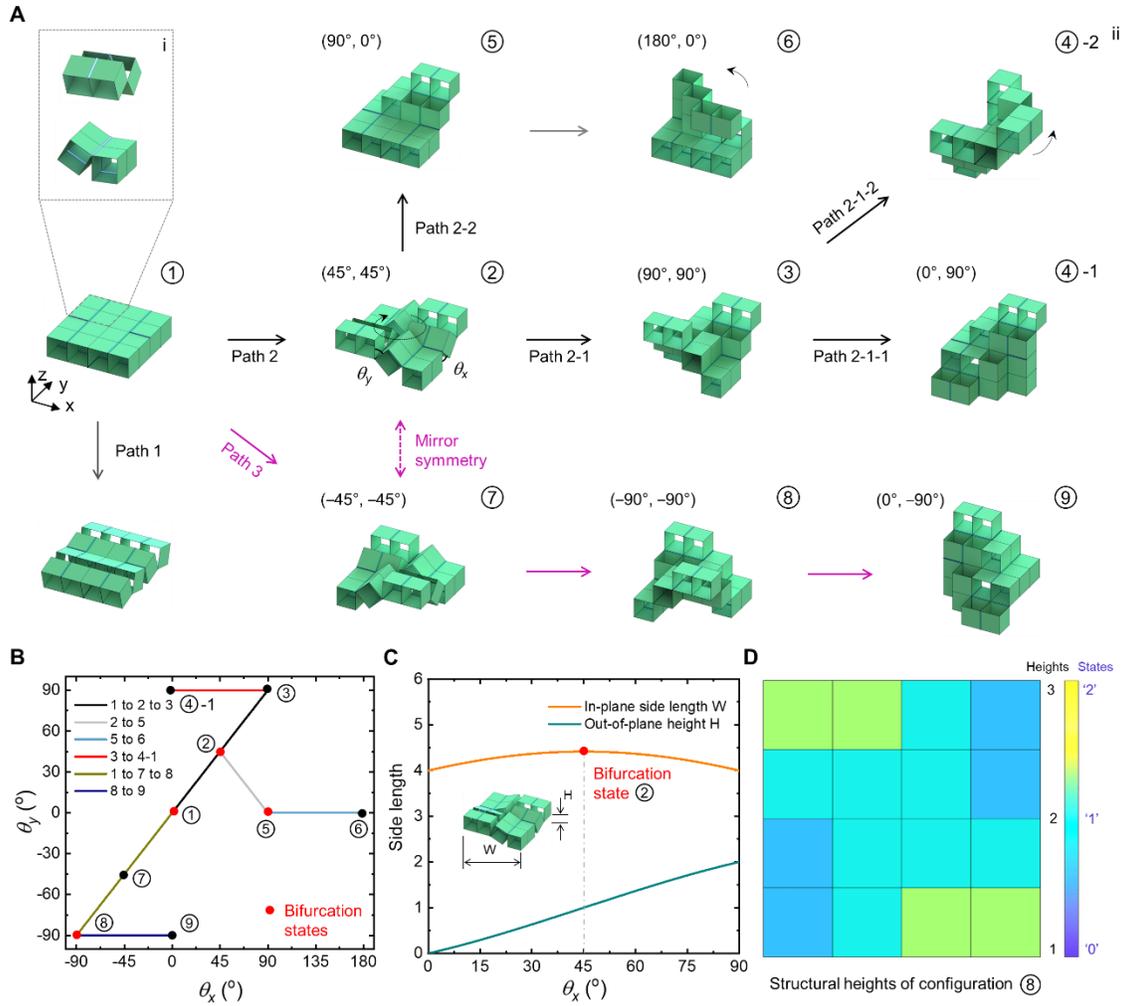

**Fig. 2. The reconfiguration details of structural unit used for constructing building block for mechanical metastructure.** (**A**) Two different reconfiguration processes of the structural unit based on the basic linkage like deformation of its composed level-1 structure, i.e., Path 1 reconfigured as linkage structure and Path 2 and 3 with novel reconfigured with structural loop accompanying with kinematic bifurcated configurations ② (in Path 2) and ⑦ (in Path 3) (ii). (**B**) The relations of the angle defined (i.e., $\theta_x$ and $\theta_y$, see details in configuration #2 in A(ii)) to describe the reconfigured configurations ① to ⑨ in A(ii). (**C**) The symmetrical deformation features of the structural unit reflected on the variation of its side length W (see the definition in the inset) with respect to the opening angle $\theta_x$ increasing from 0° to 90°, and the kinematic bifurcation point occurs at $\theta_x = 45°$. (**D**) Illustration of the non-overlap structural feature of the post-bifurcated configuration from Path 3 (i.e., configuration ⑧) though the contour map of structural height when reaching to compact state.

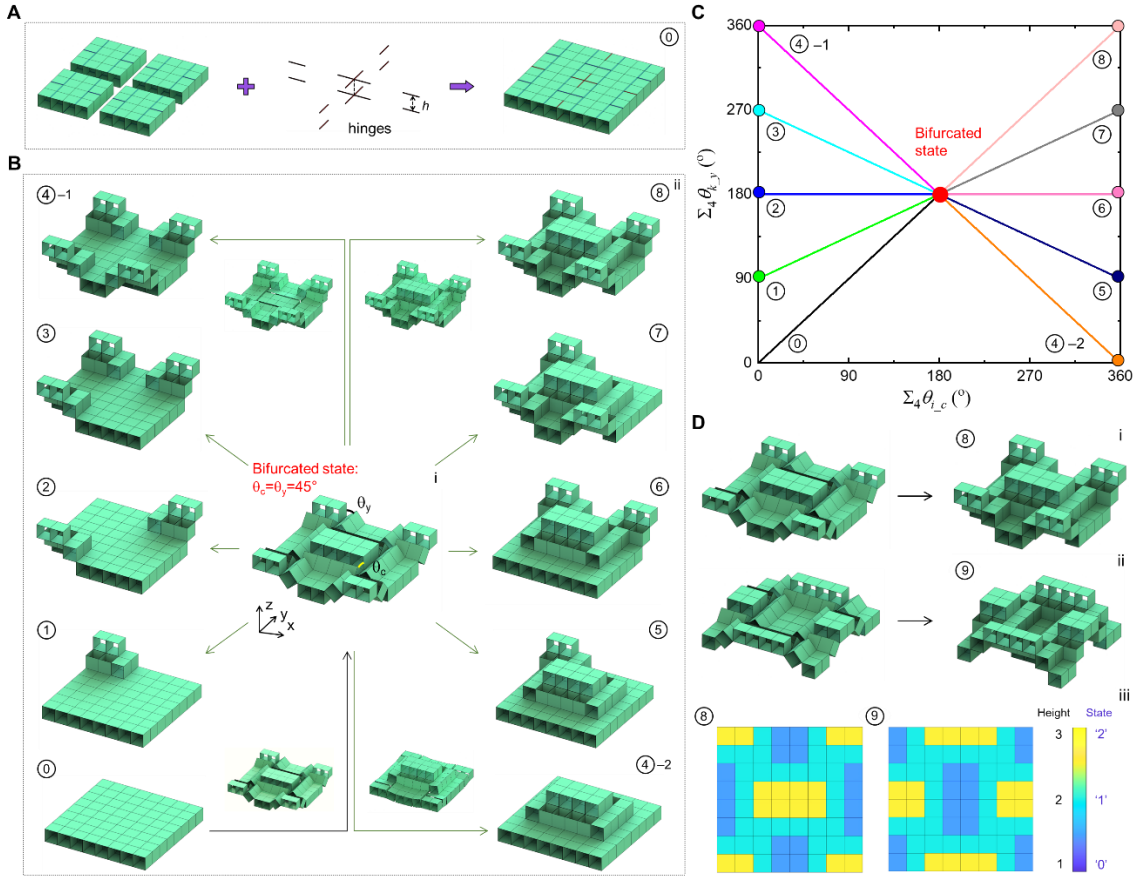

**Fig. 3. Schematics and reconfigurations of the building block.** (**A**) The construction of building block by connection four structural units with 16 line hinges with spatial arrangement shown in the middle. (**B**) Reconfigurations of the building block with kinematic bifurcations: starting from the initial configuration firstly to the bifurcated configuration ① with the defined corner and side opening angle equal to 45° and five independently deformable structural segments (i.e., one middle one surrounded by four cornered segments) by whose combinatorics the following reconfigured configuration branches can be obtained, see the representative configurations ① to configuration ④-1 by popping up only the corner segments and the other five representative configuration ④-2 to ⑧. (**C**) The angle relations to describe each configurations in (B). (D) Illustration of the non-overlap structural feature of the post-bifurcation reconfigured configurations with compact structural form: the configurations (i and ii) and the contour map of their structural heights (iii).

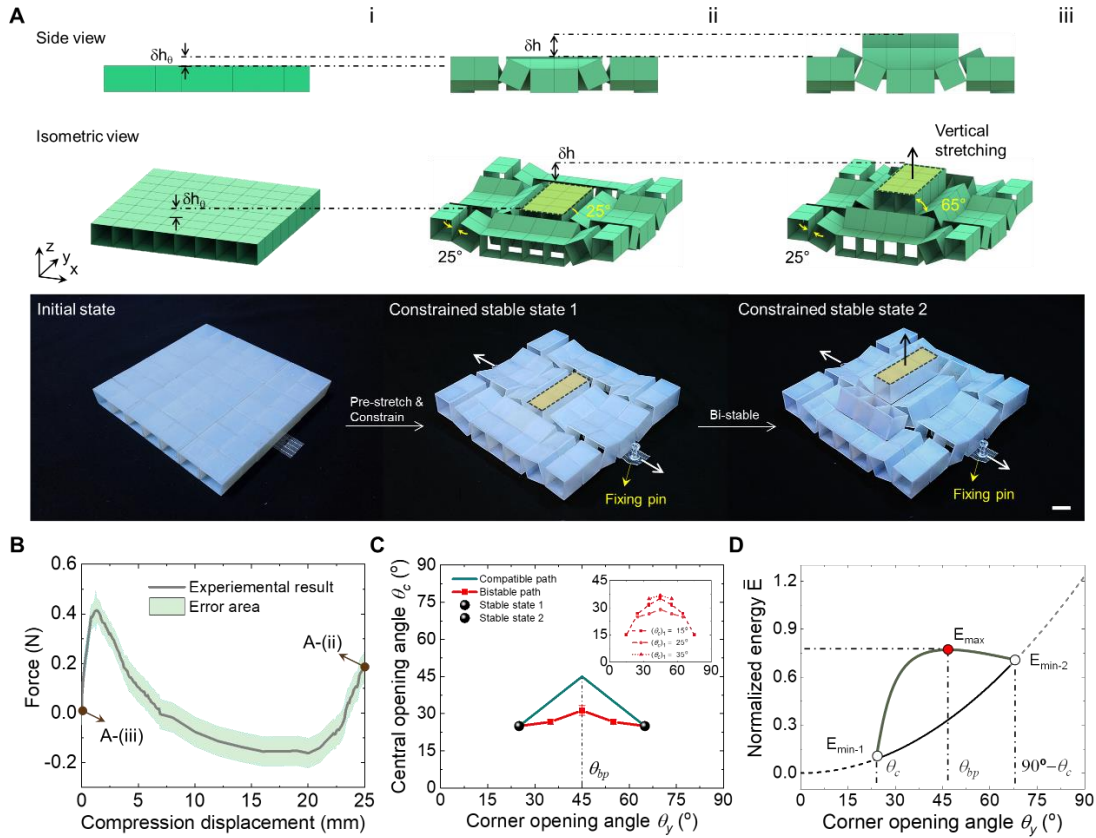

**Fig. 4. Bistable deformation demonstration of the building block with pre-stretching and boundary constraints**. (**A**) Schematic and experimental illustration to achieve bistable deformation for the building block, i.e. pre-stretching (i), constraining the boundaries (ii) and the occurrence of bistable deformation by popping up the central segment, see the highlighted part in (ii) and (iii). (**B**) Experimental demonstration of the bistable deformation of the pre-stretched and boundary constrained building block by displacement controlled uniaxial compression test. (**C**) Experimental illustration of the incompatible deformation underlying mechanism for the occurrence of the bistable deformation of building block. (**D**) Illustration of the bistable deformations of the building block through the energy curve during deformation with two energy minimum $E_{min-1}$ and $E_{min-2}$ and the energy maximum $E_{max}$ as barrier.

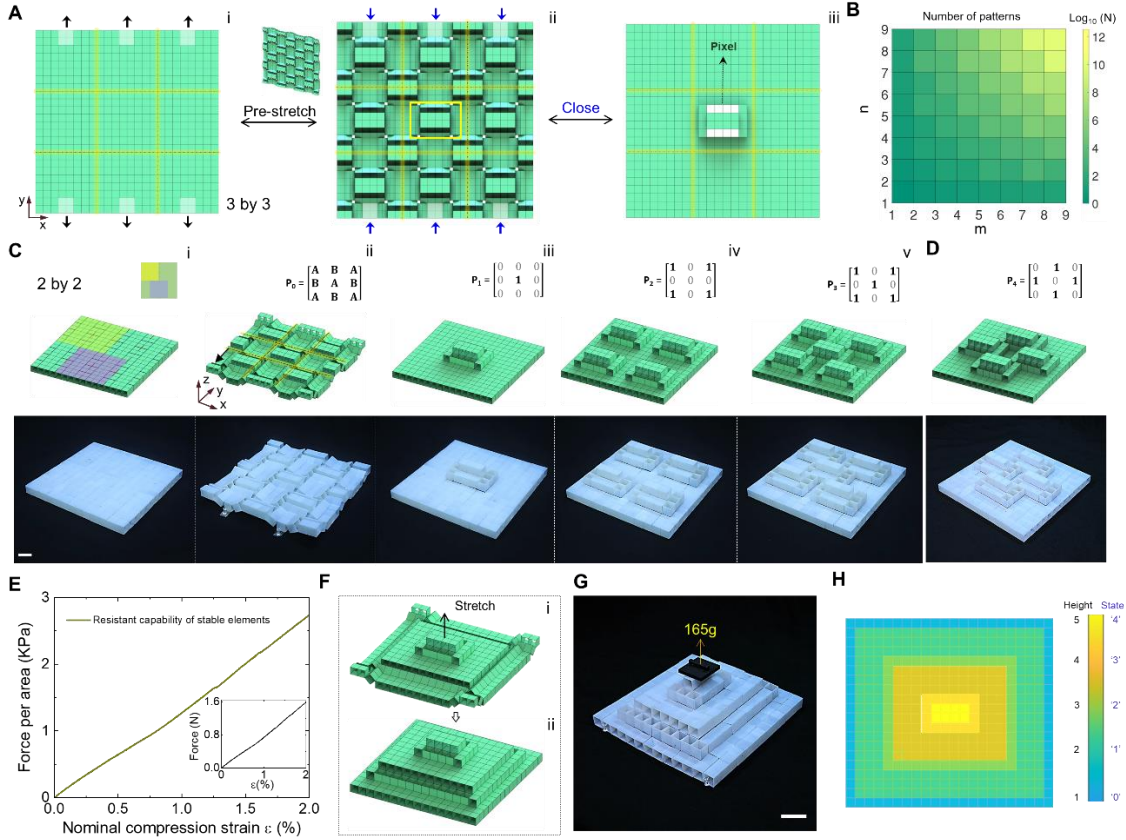

**Fig. 5. Illustration of the reprogrammable information storage features and the underlying mechanism of high information density and structural stability of the proposed mechanical metastructure**. (**A**) Schematics of the formation of one information bit as display pixel on a 3 by 3 structure, i.e., pre-stretching the initial configuration (i) to the configuration prior the bifurcated configuration (ii), then selectively popping up the central structural segment and finally releasing the pre-stretching condition form the information bit which can be the pixel for information display use (iii). (**B**) Illustration of the highly reprogrammable reconfigured information patterns of our designed mechanical metastructure related to the number of building block (m, n separately representing the number of building block along its two different sides). (**C-D**) Detailed demonstration of the reprogrammed structural patterns (ii to v, and D) based on a 2 by 2 structure (i) and one of the underlying mechanism for the high information density by generating an additional structural segments (i.e., the central one, iii). (**E**) Experimental demonstration of the high structural stability of our proposed structure after storing with certain information though uniaxial compression test (close to 2.7KPa) (the inset: tolerated external forces with respect to compression strain) (**F** to **H**) The other underlying mechanisms of the high information density features of our proposed mechanical metastructure through linearly reconfiguring into multilayered configuration with non-overlap structural features.

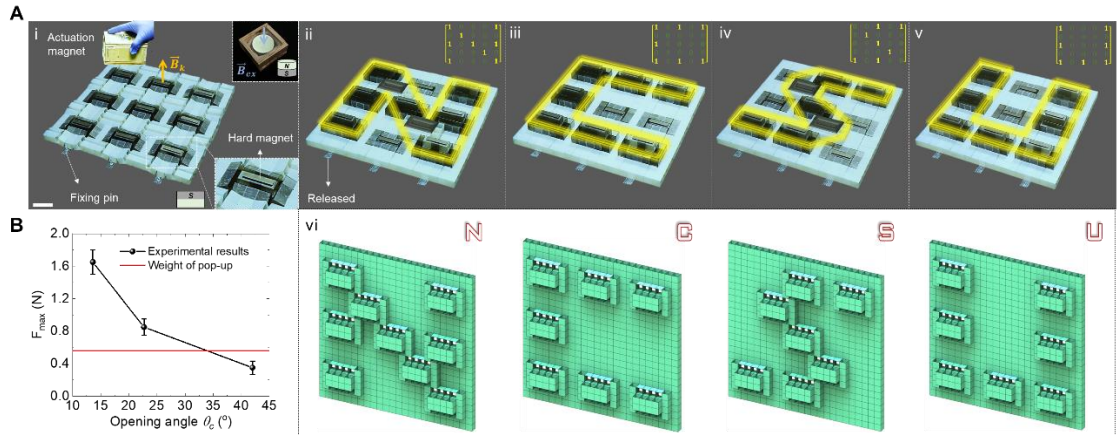

**Fig. 6. Illustrations of the multifunctional applications of our proposed mechanical metastructure.** (**A** and **B**) Mechanical information storage and interaction actuated through magnetic poles: schematic and experimental demonstrations by four letters "N", "C", "S", "U" in (A), and the affective pre-rotation angle for the successful achievements information storage by using magnetic poles (B).

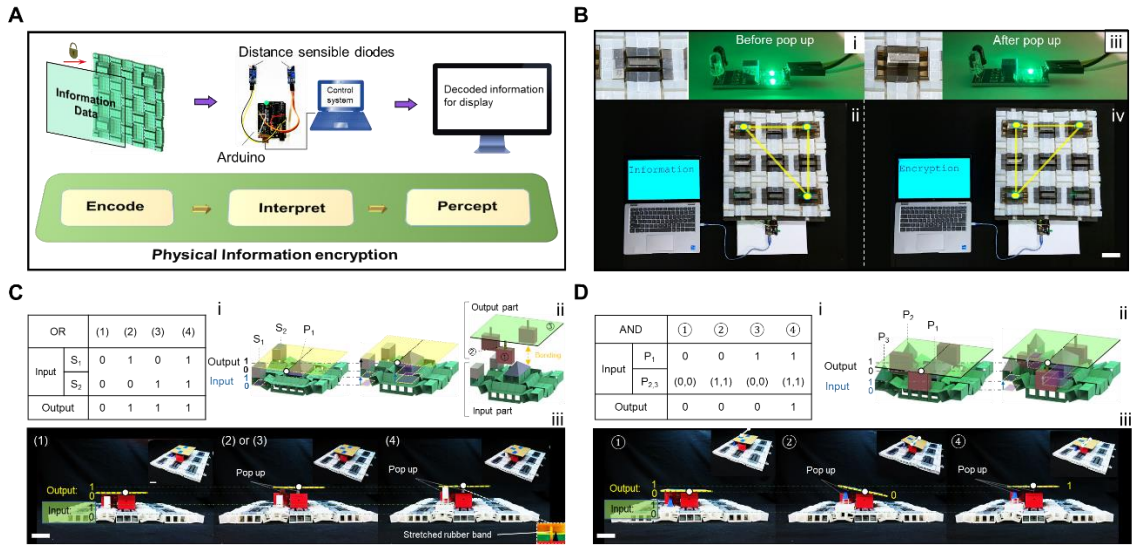

**Fig. 7. Illustrations of the multifunctional applications for information encryption and mechanical computing of our proposed mechanical metastructure.** (**A** and **B**) Mechanical information encryption (A) used through the reconfigured triangle shapes (B). (A) gives the information encryption process by firstly encoding the desired information to the reconfigured mechanical metastructure, then intercepting the encoded information through distance sensing system (i and iii, B) and finally with percept with visual display through control system (ii and iv, see the physical triangle configuration encoded two "Information" and "Encryption" letters). (**C**-**D**) Illustration of the mechanical computation uses for "OR" (C) and "AND" (D) binary logic gate uses with three-dimensional structural forms.